%% file: main.tex
\documentclass{article}

     \PassOptionsToPackage{numbers, compress}{natbib}

\usepackage[preprint]{neurips_2024}

\usepackage{algorithm}
\usepackage[noend]{algpseudocode}
\usepackage[utf8]{inputenc} 
\usepackage[T1]{fontenc}    
\usepackage{hyperref}       
\usepackage{url}            
\usepackage{booktabs}       
\usepackage{amsfonts}       
\usepackage{nicefrac}       
\usepackage{microtype}      
\usepackage[dvipsnames]{xcolor}         
\usepackage{graphicx}
\usepackage{doi}
\usepackage{subcaption}
\usepackage{array}
\usepackage{adjustbox}
\usepackage{multirow}
\usepackage{amsmath}
\usepackage{pifont}
\usepackage[flushleft]{threeparttable}
\usepackage{arydshln}
\usepackage{bm}
\usepackage{wrapfig}
\usepackage{booktabs}
\newcommand{\cmark}{\color{OliveGreen}\ding{51}}%
\newcommand{\xmark}{\color{Red}\ding{55}}%

\newcommand{\nmark}{\color{Orange}$\bm{\sim}$}
\usepackage{diagbox}
\usepackage{placeins}

\usepackage{xr}
\makeatletter
\newcommand*{\addFileDependency}[1]{
\typeout{(#1)}
\@addtofilelist{#1}
\IfFileExists{#1}{}{\typeout{No file #1.}}
}\makeatother

\newcommand*{\myexternaldocument}[1]{%
\externaldocument{#1}%
\addFileDependency{#1.tex}%
\addFileDependency{#1.aux}%
}

\myexternaldocument{supplemental}

\title{Implicit Delta Learning of High Fidelity Neural Network Potentials}
\date{May 2024}
\author{%
  Stephan Thaler \thanks{These authors contributed equally to this work.} \\
  Valence Labs\\
    \And 
    Cristian Gabellini \footnotemark[1]\\
  Valence Labs\\
  \And 
  Nikhil Shenoy \\
  Valence Labs\\
  \And
  Prudencio Tossou\thanks{Reachout at: prudencio at valencelabs.com} \\
  Valence Labs\\
}

\begin{document}

\hypersetup{
pdftitle={ImplicitDeltaLearning},
pdfsubject={cs.LG},
pdfauthor={},
pdfkeywords={}
}
\maketitle

\begin{abstract}
Neural network potentials (NNPs) offer a fast and accurate alternative to ab-initio methods for molecular dynamics (MD) simulations but are hindered by the high cost of training data from high-fidelity Quantum Mechanics (QM) methods. Our work introduces the \textit{Implicit Delta Learning }(IDLe) method, which reduces the need for high-fidelity QM data by leveraging cheaper semi-empirical QM computations without compromising NNP accuracy or inference cost. IDLe employs an end-to-end multi-task architecture with fidelity-specific heads that decode energies based on a shared latent representation of the input atomistic system. In various settings, IDLe achieves the same accuracy as single high-fidelity baselines while using up to 50x less high-fidelity data. This result could significantly reduce data generation cost and consequently enhance accuracy and generalization, and expand chemical coverage for NNPs, advancing MD simulations for material science and drug discovery. Additionally, we provide a novel set of 11 million semi-empirical QM calculations to support future multi-fidelity NNP modeling.
\end{abstract}

\section{Introduction}
\label{sec:intro}
\input{sections/introduction}
\section{Related works}
\label{sec:litterature}
\input{sections/related-works}

\section{Methods}
\label{sec:implicitdeltal}
\input{sections/implicitdeltal}


\section{Experiments}
\label{sec:experiments}
\input{sections/experiments}

\section{Discussion}
\label{sec:discussion}
\input{sections/discussion}


\bibliographystyle{unsrtnat}
\bibliography{references} 

\clearpage
\appendix
\section{Appendix}
\input{sections/additional-files}

\end{document}


\newpage 
\setcounter{page}{1} 
\renewcommand{\thepage}{S\arabic{page}} 

\title{%
\huge Supplemental Material\\[1em]
\LARGE \emph{Implicit Delta learning} is all you need\\
}

\author{
    \begin{tabular}{c@{\hskip 2em}c@{\hskip 2em}c} 
        Cristian Gabellini \thanks{Equal contribution from all authors} & Stephan Thaler & Prudencio Tossou \\
        \textit{Valence Labs}  & \textit{Valence Labs}  & \textit{Valence Labs} 
    \end{tabular}
}

\date{} 
\maketitle


\renewcommand{\thefigure}{S\arabic{figure}}
\setcounter{figure}{0} 

\renewcommand{\thetable}{S\arabic{table}}
\setcounter{table}{0} 

\label{sec:appendix}
\section{Additional figures}

\clearpage
\bibliographystyle{unsrtnat}
\bibliography{references}

%% file: sections/introduction.tex
Molecular dynamics (MD) simulations are a major computational workhorse in material science and drug discovery, enabling the in-silico study of molecular systems and their dynamics. The accuracy of an MD simulation is determined by the potential energy function, which defines intra- and intermolecular forces. \textit{Ab-initio} MD simulations employ forces computed from first-principles quantum mechanics (QM) methods such as density functional theory (DFT). While highly accurate, the unfavorably scaling cost of QM computations hinders the application of \textit{ab-initio} MD simulations for large systems or long timescales.

In recent years, neural network potentials (NNPs) \cite{Behler2007, Gilmer2017, Ko2021, Batzner2022, pelaez2024torchmd} have enabled orders of magnitude faster MD simulations with \textit{ab-initio} accuracy \cite{musaelian2023scaling, kovacs2023mace}. However, practical issues prevent their widespread development and adoption. The first issue is the large amount of high-fidelity (HF) QM data needed for training. Collecting this data is expensive and limited in terms of chemical coverage and size, resulting in fewer NNPs being developed. The second issue is the limited generalization capability of NNPs. Due to data limitations, NNPs often fail to generalize to unseen parts of chemical space \cite{Fu2022, Thaler2022force, shen2018molecular}. To improve the stability and accuracy of NNP-based MD simulations, prior potentials \cite{Wang2019, Thaler2022rel, musaelian2023scaling}, MD observable-based training \cite{Thaler2021, raja2024stability}, and/or NNP fine-tuning with additional QM data \cite{shen2018molecular, Smith2018, van2023hyperactive} are often required. To enhance the development and adoption of NNPs, it is imperative to lower the data generation costs for their initial training and fine-tuning.

One approach to reduce NNP training and fine-tuning costs is to leverage lower-fidelity (LF) methods such as Tight-Binding (TB) (e.g., GFN2-xTB \cite{bannwarth2019gfn2}, DFTB3\cite{gaus2011dftb3}) or Hartree-Fock-based Semi-Empirical (SE) methods (e.g., PM6 \cite{stewart2007optimization}). These methods are significantly less expensive, though less accurate and generalizable, compared to high-fidelity (HF) computations such as DFT or coupled cluster. The aim is to increase NNP stability, generalization, and chemical space coverage without sacrificing accuracy or requiring additional fine-tuning. $\Delta$-learning methods train ML models to predict the energy difference between an LF and a HF method \cite{ramakrishnan2015big, shen2016multiscale, shen2018molecular, qiao2020orbnet}. The underlying idea is that learning the difference $\Delta$ is more data efficient than directly predicting the HF energy. However, while this reduces the number of HF labels needed during training, it significantly increases inference cost due to on-the-fly LF QM calculations during MD simulations.

In this work, we propose the \textit{Implicit Delta Learning} (IDLe) training strategy to lower the training cost of NNPs and increase their data efficiency, chemical coverage, and generalization without requiring additional QM computations at inference. Based on a multi-task network, IDLe reduces the number of HF labels by replacing them with LF labels (typically SE labels) and implicitly learning to decode the HF energy from a shared latent representation of the input system. Our contributions can be summarized as follows:
\begin{enumerate}
    \item We propose the \textit{IDLe} training strategy to replace the majority of HF training data required by HF NNPs with LF data. Consequently, IDLe can reduce the cost of NNP training data generation by orders of magnitude, even requiring no additional HF labels in certain cases.
    
    \item We showcase IDLe's generalization effectiveness within and outside the training distribution for NNPs trained on various QM datasets relevant to drug discovery (QMugs \cite{isertQmugs}, Spice \cite{eastman2022spice, eastman_2024_10975225}, QM7-X \cite{HojaQM7X}, ANI1-ccx \cite{ani1ccx}), spanning multiple combinations of HF and LF methods.
    
    \item We performed $\sim$11 million single point energy computations using LF QM methods to augment existing datasets and demonstrate the power of IDLe. These new computations will be publicly available to foster further research into multi-fidelity NNP training and IDLe.
\end{enumerate}

%% file: sections/related-works.tex
\paragraph{$\Delta$-learning:} 
Given that LF methods capture many interaction terms of HF methods, "explicit" $\Delta$-learning aims to predict the remaining $\Delta$ using ML models. During training, samples are labeled with both LF and HF methods, and during inference, LF labels are computed and added to the predicted $\Delta$s to obtain the HF energy. Since $\Delta$s are easier to learn, $\Delta$-models are data-efficient and highly accurate during inference at a cost comparable to that of the LF computation.

The concept of $\Delta$-learning in quantum chemistry was introduced by \citet{ramakrishnan2015big}, who demonstrated its data efficiency compared to direct HF learning. Their $\Delta$-models learned and predicted differences in levels of theory (e.g., PM6 $\rightarrow$ CCSD(T)), across quantum properties (e.g., energy $\rightarrow$ enthalpy), and geometries. Later, \citet{boselt2021machine} and \citet{hofstetter2022graph} applied $\Delta$-learning to incorporate long-range interactions in their NNPs for condensed-phase and reactive systems by learning the energy differences between DFT and SE methods. DelFTa \cite{isertQmugs} predicts a wide array of quantum properties by $\Delta$-learning from GFN2-xTB to $\omega$B97X-D/def2-SVP.

\paragraph{High-fidelity NNPs at low cost:} While $\Delta$-learning reduces training data costs, its inference cost is high for most applications. Consequently, alternatives such as active learning \cite{Smith2018, Zhang2019, Jinnouchi2020}, transfer learning \cite{Smith2019}, meta-learning \cite{allen2023learning}, and multi-task pre-training \cite{shoghi2023molecules} aim to minimize the amount of HF training data
while avoiding increased inference cost.

Active learning (AL) collects the most informative HF data based on the NNP predictive uncertainty, significantly reducing training data size \cite{Smith2018, Zhang2019, Jinnouchi2020}. For example, ANI-1x \cite{comp6_ani1x, ani1x_ani1ccx}, built with AL, has $\sim$4x less DFT data than ANI-1 \cite{ani1}, yet training NNPs on ANI-1x yields superior performance.

Transfer learning methods use cheap LF data to pre-train an NNP and fewer HF calculations for fine-tuning. For example, \citet{Smith2019} pre-trained an NNP on $\sim5$ million DFT points before fine-tuning with 0.5 million CCSD(T)* samples.
Similar transfer learning applications typically transfer representations within the same part of chemical space, transferring from rather costly methods such as DFT or Hartree-Fock to higher-fidelity methods as the target \cite{tsubaki2021quantum, kolluru2022transfer, zaverkin2023transfer, chen2022machine, kaser2023transfer}. However, transfer learning NNPs based on cheap SE methods has not yet been attempted, despite very recent applications in molecular property prediction \cite{buterez2024transfer, kirschbaum2024transfer} and earlier work using "synthetic data" \cite{gardner2024synthetic} .
Transfer learning has two stages, is not end-to-end, and can require extensive training computations if both the LF and the HF data are sizable. Selecting the pre-training dataset to transfer from for a given HF target can also be daunting.

Meta-learning \cite{allen2023learning} and multi-task networks \cite{shoghi2023molecules} can also be used for pre-training, learning from different parts of chemical space using multiple QM datasets before fine-tuning to a specific HF QM method or domain of chemical space. They share the same limitations as transfer learning methods.

\paragraph{Multi-fidelity QM datasets:} Numerous QM datasets have been created for NNP training, but only a few include multi-fidelity QM labels that improve training efficiency for HF NNPs. The largest is PubChemQC \cite{nakata2023pubchemqc}, which contains DFT (B3LYP/6-31G*) and SE (PM6) properties for $\sim86$ million geometries covering a wide range of molecules. Similarly, QMugs \cite{isertQmugs} includes $\sim2$ million conformations from 665,000 molecules with up to 100 heavy atoms, providing DFT ($\omega$B97M-D/def2-SVP) and TB (GFN2-xTB) energies and properties. The QM7-X dataset \cite{HojaQM7X} and MultiXC-QM9 \cite{nandi2023multixc} cover smaller molecules with restricted chemical diversity. QM7-X contains DFT (PBE+MBD) and TB (DFTB3+) energy labels for $\sim4$ million conformations of molecules with up to seven heavy atoms. MultiXC-QM9 provides multi-fidelity data with 72 DFT functionals and a TB method (GFN2-xTB). None of these datasets overlap in chemical space or feature the same HF QM method pairs, limiting the study of extrapolation in multi-fidelity or $\Delta$-learning NNPs.

%% file: sections/implicitdeltal.tex
\subsection{Preliminaries}
NNPs can be trained by minimizing an energy-matching mean squared error (MSE) loss function \cite{Behler2007}. For high-fidelity NNPs, we have
\begin{equation}
    \mathcal{L} = \frac{1}{N} \sum_{i=1}^N \left[E_i^\mathrm{HF} - \hat{E}_{\bm\theta}^\mathrm{HF}(\mathbf{S}_i)\right]^2 \ ,
\end{equation}
where $\mathbf{S}_i$ is a conformation of a molecule with $K_i$ atoms, typically represented by positions $\mathbf{R}_i \in \mathbb{R}^{K_i \times 3}$ and atomic numbers $\mathbf{Z}_i \in \mathbb{N}^{K_i}$, $E_i^\mathrm{HF}$ is the target HF energy, and $    \hat{E}_{\bm\theta}^\mathrm{HF}(\mathbf{S}_i)$ is the predicted energy of the NNP parameterized by $\bm \theta$. \\
Instead of directly learning and predicting the HF energy $\hat{E}_{\bm\theta}^\mathrm{HF}(\mathbf{S}_i)$, the $\Delta$-learning method \cite{ramakrishnan2015big} learns to predict the energy difference with respect to a LF energy $E^\mathrm{LF}$. At inference, predicting the HF energy requires the computation of $E^\mathrm{LF}$ and the $\Delta$, i.e.,
$
    \hat{E}_{\bm\theta}^\mathrm{HF}(\mathbf{S}_i) = E^\mathrm{LF}_i + \Delta \hat{E}_{\bm\theta}^\mathrm{HF-LF}(\mathbf{S}_i)$.
The parameters of the $\Delta$-model are learned by minimizing the MSE loss function using samples where we have both the HF and LF labels as
$
    \mathcal{L}^{\Delta} = \frac{1}{N} \sum_{i=1}^N \left[(E_i^\mathrm{HF} - E_{i}^\mathrm{LF}) - \Delta E_{\bm\theta}^\mathrm{NN}(\mathbf{S}_i)\right]^2
$.
Due to the form of this loss, we need pairs of HF and LF energies for any given geometry in our training data, which makes data collection quite expensive. Additionally, $\Delta$-learning methods are straightforward to apply for datasets with one LF and one HF method, but generalizing them to multiple LF methods and a HF one seems challenging.

\subsection{Implicit Delta Learning}

To mitigate the limitations of existing learning approaches when leveraging LF data to improve the data efficiency and accuracy of HF NNPs, we propose the Implicit Delta Learning training approach, named IDLe. In comparison to $\Delta$-learning, IDLe can generalize easily to multiple low-fidelity methods and does not need LF computations to be very accurate during inference. 
Relative to transfer learning methods, it is an end-to-end approach that leverages LF and HF data simultaneously instead of the two stages of pre-training and fine-tuning to allow transfer. The full comparison between IDLe and other training approaches for HF NNPs is shown in Table~\ref{tab:comp_qualitative}.

Technically, IDLe is a multi-task learning approach, where each prediction head corresponds to a different QM method. All prediction heads share the same backbone network, forcing them to decode their respective energy value from a shared latent representation of the input geometry. 
Due to this framing, one can leverage multiple LF labels to improve HF prediction accuracy.

In order to train a model via IDLe, we minimize the following multi-fidelity multi-task MSE loss function:
\begin{equation}
    \mathcal{L}^{MT} = 
    \frac{1}{M} 
    \sum_{i=1}^N \sum_{h=1}^{H} I_{i, h} \times \left[E_{i,h} - \hat{E}_{\bm\theta, h}(\mathbf{S}_i)\right]^2 \ ,
\end{equation}
where $M$ is the number of energy labels in the dataset, $N$ is the number of geometries, $H$ is the number of QM methods (both LF and HF) considered, $I_{i, h}$ is 0 if the label $E_{i,h}$ is missing, otherwise 1, and $\hat{E}_{\bm\theta, h}(\mathbf{S}_i)$ is the energy prediction of head $h$ for the geometry $\mathbf{S}_i$.

\begin{table}[ht]
\caption{Desiderata fulfillment of different high fidelity NNP training methods.}
\label{tab:comp_qualitative}
\centering
\resizebox{\textwidth}{!}{\begin{tabular}{lccccc}
\toprule
 & Direct-learning & $\Delta$-learning & Active learning & Transfer learning  &  IDLe    \\ 
 \midrule
\textit{Training data efficiency} & \xmark & \nmark & \cmark & \cmark & \cmark  \\ 
\textit{Inference cost} & \cmark & \xmark & \cmark & \cmark & \cmark  \\ 
\textit{Accuracy} & \cmark & \cmark\cmark & \cmark 
 & \cmark& \cmark  \\ 
\textit{Leverage LF labels} & \xmark & \nmark & \xmark & \cmark & \cmark  \\ 
\textit{OOD generalization} & \xmark & \cmark & \xmark & \nmark & \cmark  \\ 
\bottomrule
\end{tabular}}

\end{table}

As the LF energies computation costs are very small compared to the HF ones (especially for larger molecules and expensive HF methods), we assume that for all conformations with HF labels we also have the LF  ones. The reverse is not true, which leads to highly different numbers of HF and LF labels during training. The training data efficiency of IDLe compared to direct and $\Delta$-learning stems from using significantly fewer HF compared to LF labels.
Under these circumstances,
IDLe implicitly learns a shared representation from which HF and LF energies can be decoded,
enabling a better chemical coverage for the HF head due to the large number of conformers with only LF labels.
In fact, the HF head benefits from the latent representation optimized via the low-fidelity prediction heads, increasing its domain of applicability without requiring further high-fidelity labels. During inference, IDLe simply uses the HF head for predictions, which is significantly cheaper than $\Delta$-learning methods, which need to compute the LF label in addition to the $\Delta$ prediction.

\subsection{Generating Multi-Fidelity data}
To assess the benefits of multiple and varying degrees of LF data on HF-NNP training, we constructed datasets where geometries have at least one HF energy label (DFT, CCSD(T)) and two LF labels: a tight binding (TB)  and an SE based on the Hartree-Fock. TB is considered slightly higher fidelity than Hartree-Fock-based SE labels, but both often have the same computational cost for small systems, even if TB is more expensive to scale to large molecules.
We expanded existing datasets, covering several parts of chemical space, with LF QM labels, which were fast to compute. The original datasets are the following:
\begin{itemize}

    \item \textbf{ANI1-ccx} \cite{ani1ccx}: ANI1-ccx has uncharged small molecules covering four elements. The energies are calculated using the CCSD(T)/CBS level of theory. Despite the limited chemical coverage, this dataset will allow the study of the extent to which LF labels can help NNPs to predict the highest fidelity energies.

    \item \textbf{Spice} \cite{eastman2022spice}: SpiceV1, containing originally $\sim1.1$ million conformations of small molecules, dimers, dipeptides, and solvated amino acids covering 15 elements. SpiceV2 \cite{eastman_2024_10975225} extends to bulk water systems, solvated molecules, and two more elements. HF energies are calculated using the $\omega$B97M-D3(BJ)/def2-TZVPPD level of theory.  

    \item  \textbf{QM7-X} \cite{HojaQM7X} contains DFT (PBE+MBD) and TB (DFTB3+) energy labels for $\sim4$ million conformations of molecules up to seven heavy atoms.

    \item \textbf{QMugs} \cite{isertQmugs} has $\sim2$ million geometries from 665k 
    molecules with up to 100 heavy atoms and provides DFT ($\omega$B97M-D/def2-SVP) and TB (GFN2-xTB) energy labels.
\end{itemize}

Table~\ref{table:data_generation_table} summarizes the QM methods existing in each datasets and the ones we have added. Note that we suffixed each dataset name by "vL" to refer to their new version with added labels.
To ensure data consistency, all the computed datasets underwent further processing to detect potential errors from the single point calculations. Any errors identified during this stage underwent a manual recalculation and subjected to the same post-processing procedure. A few molecular structures were systematically removed after multiple rounds of re-calculation without convergence. 

All the calculations were set up using QCSubmit \cite{orthonQCSubmit} and the computations were managed with QCFractal \cite{Smith2021QCFractal}. The backends for the single point calculations are MOPAC \cite{mopacArticle} for PM6 and xTB \cite{xtbArticle} for GFN2-xTB.

\begin{table}[ht]
\caption{Summary of the generated datasets with QM methods and number of conformers.}
\label{table:data_generation_table}
\centering
\resizebox{\textwidth}{!}{
\begin{tabular}{@{}ccccccc@{}}
\toprule
\multirow{2}{*}{Dataset} & \multicolumn{2}{c}{Original labels} & \multicolumn{3}{c}{Added Labels} & \multirow{3}{*}{\# Conformers} \\ \cmidrule(lr){2-6}
      & High-fidelity QM  & TB  & DFT & TB   & SE   &     \\ 
 \midrule
 ANI-1ccxvL & CCSD(T)/CBS &  & $\omega$B97X-d/def2-svp & GFN2   & PM6 & 489,457 \\ 
 SpicevL2   & $\omega$B97M-D3(BJ)/def2-TZVPPD  &  &   & GFN2  & PM6 & 2,004,893  \\ 
 QM7-XvL    & PBE0+MBD    & DFTB3+  &  &    & PM6  & 4,195,192  \\ 
QMugsvL   & $\omega$B97X-D/def2-SVP    & GFN2  &  &  & PM6   & 1,992,941   \\ 
\bottomrule
\end{tabular}}

\end{table}

%% file: sections/experiments.tex
\subsection{Setup}
\label{sec:setup}

We perform all experiments using TorchMDNet 2.0 \cite{pelaez2024torchmd} with prediction heads consisting of MLPs with a single hidden layer; a detailed list of hyperparameters is provided in Appendix \ref{appendix:hyperparameters}. 
We compute normalized energy labels $\Tilde{E}$ for each dataset and QM level of theory $j$ by subtracting the mean energy value of each atom type as computed by a linear model $\mu_{j,k}$, analogous to \cite{ramakrishnan2015big}. In addition, we scale the energy by the mean residual energy per atom $\sigma_j$:
\begin{equation}
    \Tilde{E}^\mathrm{QM}_{i,j} = \frac{1}{\sigma_j} \left(E^\mathrm{QM}_{i,j} - \sum_{k=1}^{K_i} \mu_{j,k}\right) \ .
\end{equation}
For all datasets, we perform a random 80\%-10\%-10\% training-validation-test split. For all the experiments, we vary the amount of HF labels from 1\%, 2.5\%, 10\%, 25\% to 100\% of the combined training and validation set, but we always evaluate generalization on the full HF test set.

We compare the data efficiency of IDLe to several competing training approaches: direct learning, $\Delta$-learning, and fine-tuning (transfer learning). Direct learning is trained exclusively on the available HF data, serving as a baseline to evaluate the effectiveness of leveraging LF data through IDLe.
$\Delta$-learning is used to assess the difference between implicitly and explicitly leveraging the LF energies with respect to predicting the HF labels.
Pre-training on SE labels and fine-tuning on the HF data will use the same amount of labels as IDLe, and therefore a comparison will show if any benefits with IDLe are due to the larger amounts of labels or the end-to-end training leading to better latent representations.

\subsection{DFT data efficiency}
\label{sec:iid_dft}
Our first experiment aims to measure IDLe's effectiveness at leveraging LF labels and compare it with other baselines for within distribution (IID) generalization. Herein, we only use QMugsvL and QM7-XvL and leverage all available LF data while varying the amount of HF data. 

\begin{figure}[h]
        \centering
    \begin{subfigure}{0.49\textwidth}   
        \centering 
        \includegraphics[trim= 10 5 30 30,clip,width=\textwidth]{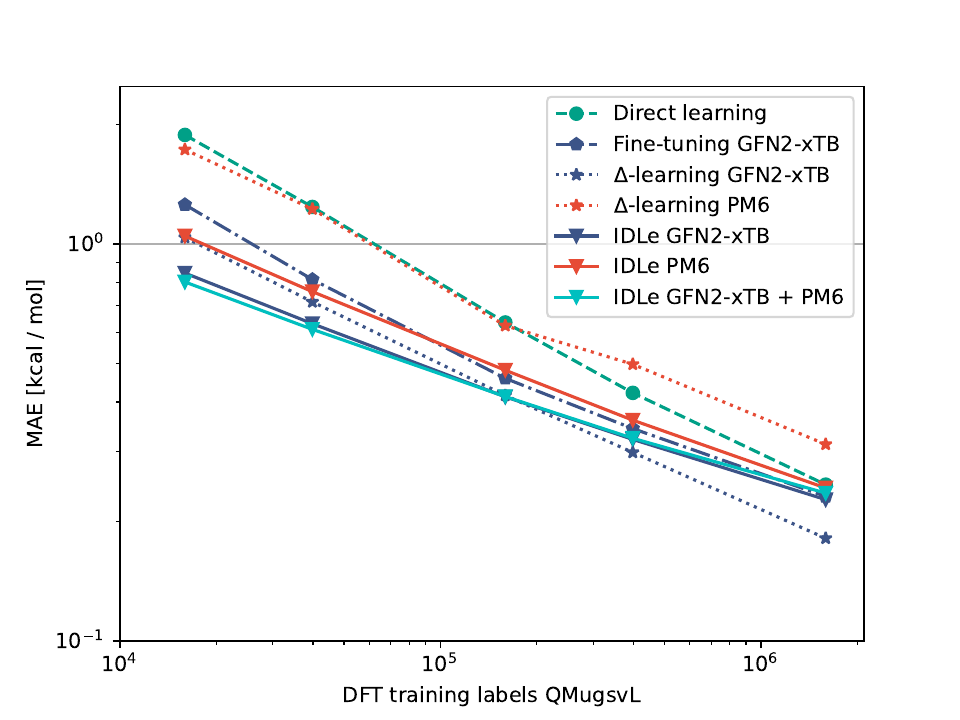}
    \end{subfigure}
    \begin{subfigure}{0.49\textwidth}   
        \centering 
        \includegraphics[trim= 10 5 30 30,clip, width=\textwidth]{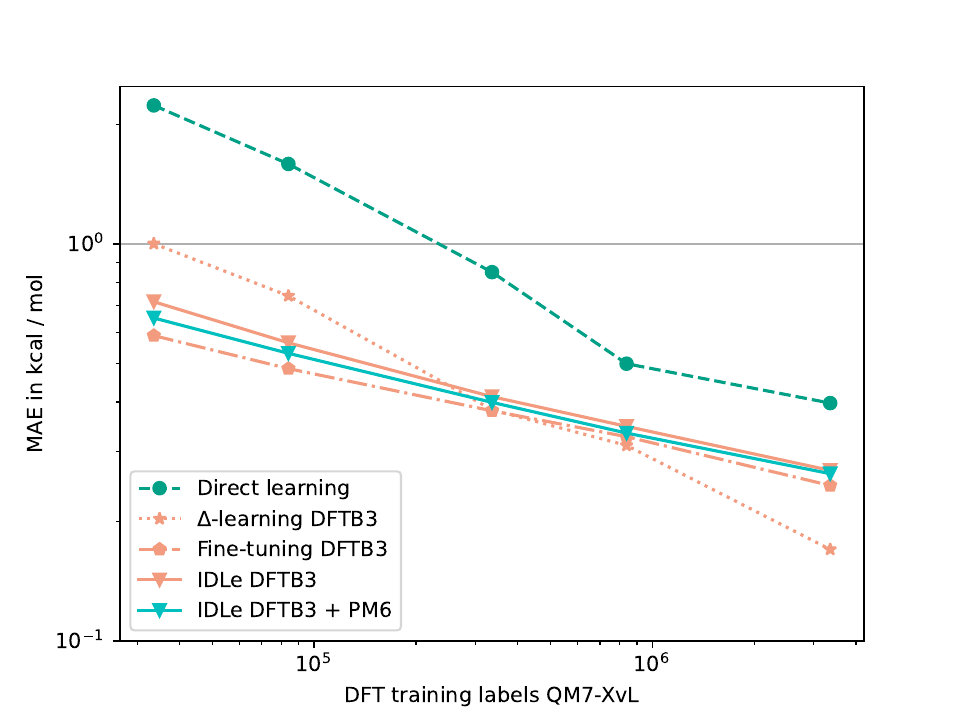}
    \end{subfigure}
    \caption{IID performance with DFT as HF labels. We compare of MAE of IDLe with several baselines for a varying amount of DFT labels on QMugs (left) and QM7-X (right).}
    
    \label{fig:iid_qmugs_qm7x}
\end{figure}

Figure~\ref{fig:iid_qmugs_qm7x} gives an overview of the results and circumstances under which SE labels help. For both datasets, at all amounts of DFT data and LF methods, direct learning significantly underperforms relative to other methods, highlighting the benefits of leveraging SE data. 
For IDLe and fine-tuning approaches, the gap relative to direct learning starts wide in low data settings and shrinks when enough HF data is available. We explain the shrinking gap by the experiment setup herein, which only measures IID generalization. In fact, when the amount of HF increases, IDLe and fine-tuning become closer to direct learning and provide less ability to leverage prior knowledge about geometries compared to direct learning. In other words, in IID settings, the quality of the representation is mostly dictated by the HF labels, but LF ones give
useful approximations of those representations in the absence of HF labels.
The gap between direct learning and $\Delta$-learning remains consistent as the number of samples increases, as expected. Due to the availability of the SE labels during inference, $\Delta$-learning becomes easier and more accurate with more samples. However, the quality of the LF labels significantly impacts $\Delta$-learning accuracy. On QMugsvL, $\Delta$-learning with PM6 is significantly less accurate than with GFN2-xTB, and this is confirmed by other experiments (see Figure \ref{fig:spice}). The quality of the LF method also affects IDLe, which is always better with GFN2-xTB relative to PM6, but to a smaller extent than $\Delta$-learning. Interestingly, using both PM6 and TB labels marginally improves the performance of IDLe. From a QM perspective, these results make sense as the energy difference between DFT methods and PM6 is significantly larger compared to TB methods (GFN2-xTB and DFTB3).

On QMugsvL, IDLe outperforms fine-tuning with a shrinking gap as the number of DFT labels increases. The opposite happens for the QM7-XvL dataset, meaning that more experiments need to be conducted to define the ideal approach in IID settings for a given pair of HF and LF methods. However, it is worth noting that on both datasets, IDLe and fine-tuning approaches reach chemical accuracy (1 kcal/mol) with 4-6x fewer DFT labels than direct learning.

\subsection{Beyond DFT data}\label{sec:iid_ccsd}
\begin{wrapfigure}{r}{0.5\textwidth}
    \centering
    \includegraphics[trim= 10 5 30 30, clip, width=0.5\textwidth]{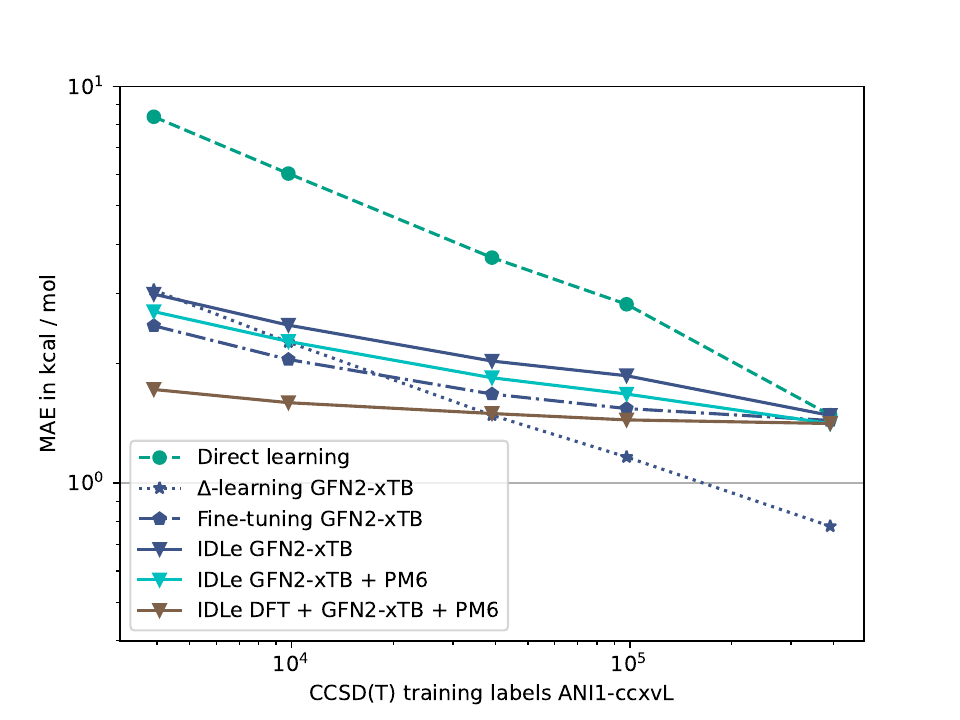}
    \caption{IID performance with CCSD(T) as HF labels: MAE of IDLE compared to several baselines for a varying amount of CCSD(T) labels on ANI1-ccxvL.}
    \label{fig:iid_ani1ccx}
\end{wrapfigure}
Learning to predict the gold-standard coupled cluster CCSD(T) level of theory based on SE or TB methods is a challenging problem \cite{allen2023learning}. This experiment seeks to address this challenge in an IID setting. Figure~\ref{fig:iid_ani1ccx} shows the generalization performance as the amount of CCSD(T) data seen by the different methods increases. It confirms the superiority of methods leveraging LF labels and their behaviors relative to direct learning, as discussed in Section~\ref{sec:iid_ccsd}.
However, herein PM6 together with GFN2-xTB significantly improves the performance of IDLe over GFN2-xTB alone. When adding DFT ($\omega$B97X-d) labels, which are LF and inexpensive compared to CCSD(T) labels, IDLe achieves almost the performance of the 100\% data CCSD(T) baseline model using only 2.5\% of CCSD(T) labels (40x data efficiency). 

Interestingly, even when using GFN2-xTB labels and 1\% CCSD(T) labels, fine-tuning and IDLe achieve better MAE than direct learning with 25\% CCSD(T) labels. This highlights the enormous advantage of using SE data to significantly reduce the number of required HF labels, even with regards to the most accurate and expensive levels of theory.
This hints further that the latent representations learned with the SE methods are truly approximative of the highest quality ones that could be learned with the CCSD(T) energies themselves.
While IDLe combines the directly-learned HF representation with the LF one, fine-tuning can override parts of the LF representation, which seems to be beneficial for GFN2-xTB in this experiment. Hence, we hypothesize that the performance difference between IDLe and transfer learning depends on the amount of mutual information (MI) between the LF and HF labels. IDLe  leverages better a high MI, while transfer learning is more robust to low MI due to its ability to unlearn LF representations that are not relevant for HF predictions.
That said, since the GFN2-xTB labels themselves contain enough information to allow $\Delta$-learning to achieve chemical accuracy in this extremely difficult setting, we hypothesize that the network capacity might be a limiting factor.

\subsection{Out-of-distribution generalization}
This section discusses the efficiency of IDLe for varying degrees of distribution shifts that we encounter in practice. For example, given two datasets $A$ and $B$ covering different parts of chemical space and sharing a HF method, a HF NNP trained on $A$ can be intended for usage on $B$ while leveraging all their LF labels. In our experiments, we train with all LF labels of $A$ and $B$, all HF labels of $A$ and vary the amount of HF labels of $B$. When this amount is zero, it means that we cannot afford any additional HF computations, so it is a "0-shot learning" setting and an extrapolation to a novel part of the chemical space. This is particularly relevant because there are many molecules for which we can only compute LF labels, but no HF ones (e.g. proteins). 

\paragraph{Chemical transferability:}
Using SpiceV1 and SpiceV1->2 (i.e. the difference between SpiceV2 and SpiceV1) as datasets $A$ and $B$ respectively, we assess the benefits of IDLe in  improving chemical transferability out-of-distribution (OOD, see Appendix~\ref{appendix:soap}). 
Figure \ref{fig:spice} (left) shows the average MAE when excluding the PubChem-Boron-Silicon subset of SpiceV1->2, which contains novel chemical elements (B, Si). Without any additional fine-tuning, the direct learning model performs rather poorly compared to all other methods leveraging LF labels.
$\Delta$-learning and fine-tuning achieve good 0-shot accuracy, but are outperformed by IDLe using only TB labels. IDLe with TB+ SE labels performs best, achieving the same accuracy as the direct learning baseline with approximately 50\% HF labels. Using only 2.5\% HF labels and TB+SE labels, IDLe achieves the same accuracy as 100\% of Spice1->2, suggesting that the cost associated with expanding SpiceV1 into SpiceV2 could have been reduced significantly by labeling ~40x fewer geometries with DFT. 

\begin{figure}[h]
    \centering
    \makebox[\linewidth][c]{
    \begin{subfigure}{0.45\textwidth}   
        \centering 
        \includegraphics[trim= 10 5 30 30, clip,width=\textwidth]{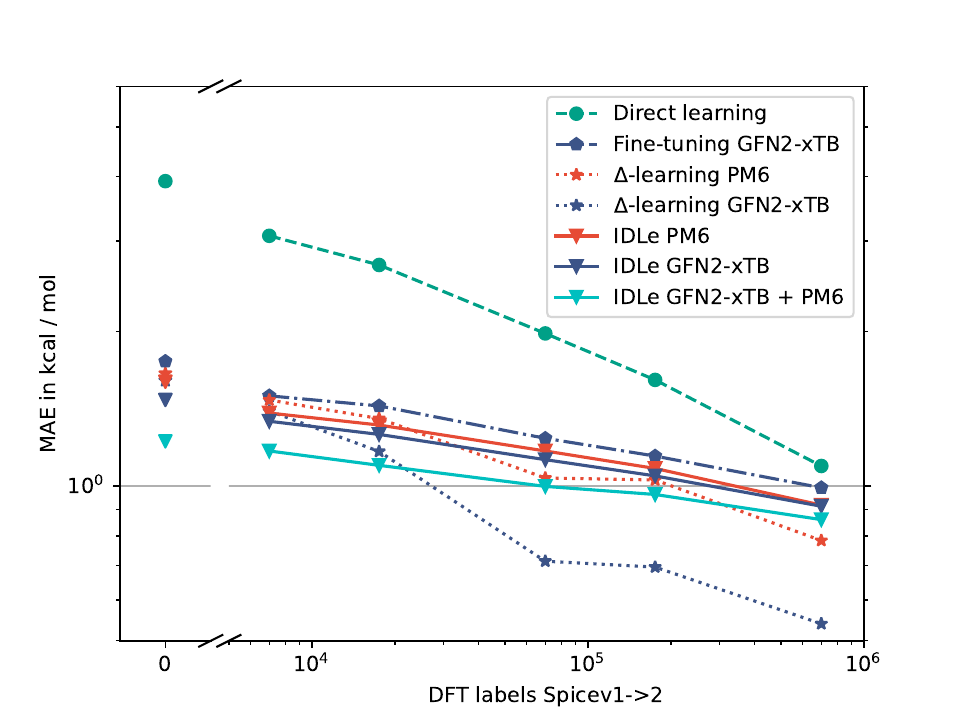}
    \end{subfigure}
    \begin{subfigure}{0.7\textwidth}   
        \centering 
        \includegraphics[width=\textwidth]{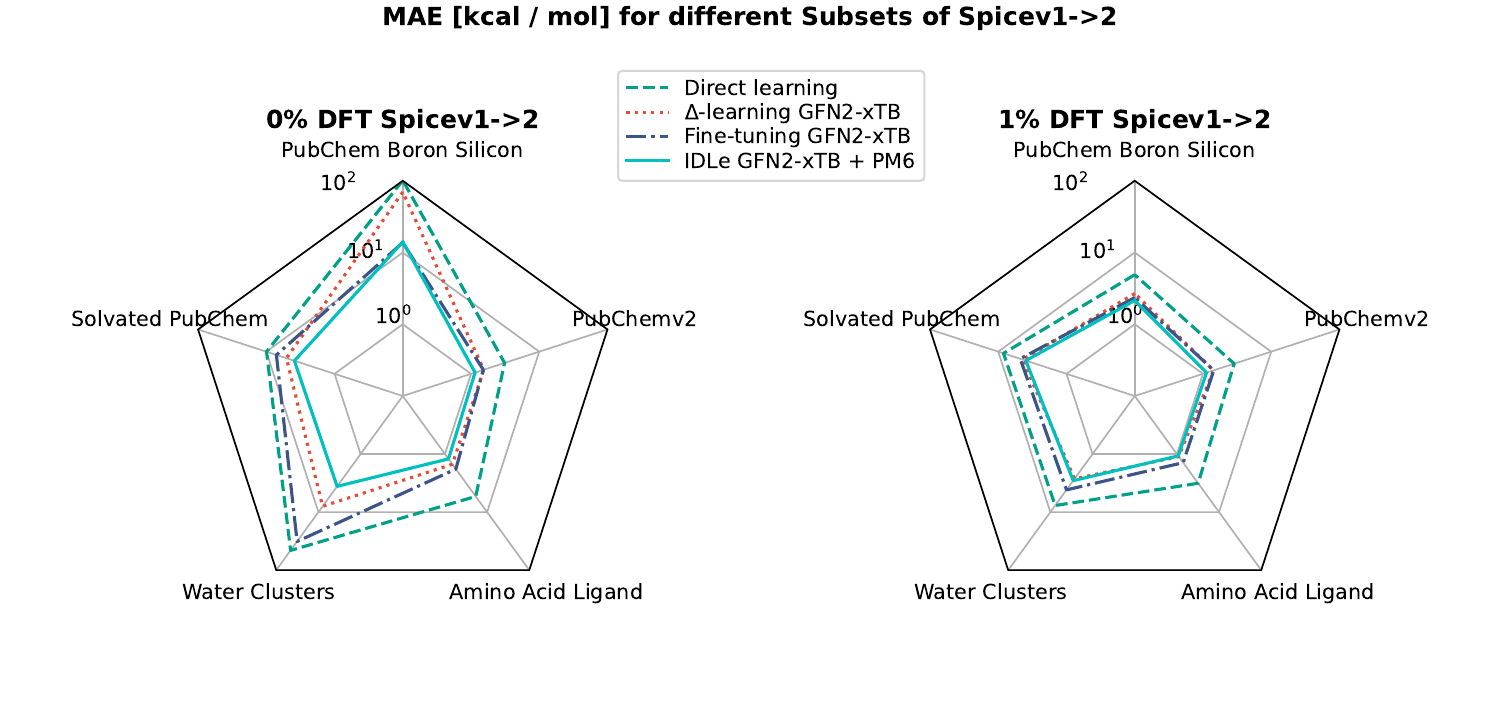}
    \end{subfigure}
    }

    \caption{OOD performance on SpiceV2->1 with $\omega$B97M-D3(BJ)/def2-TZVPPD as HF labels: On the left, we show the average MAE of IDLe compared to other methods for a varying amount of DFT labels (without the PubChem-Boron-Silicon subset). On the right, we distinguish between all the subsets of SpiceV1->2 with $0\%$ and $1\%$ of the HF labels available during training.}
    \label{fig:spice}
\end{figure}

The MAE on the individual subsets of SpiceV1->2 is shown in Figure~\ref{fig:spice} (right). For all subsets, in both the 0-shot and 1\% DFT label scenarios, IDLe outperforms all the other methods. For the subsets with small to medium distribution shifts (PubChemV2 and Water Clusters), IDLe with TB + SE labels in the 0-shot setting achieves the same accuracy as direct learning with 100\% HF labels, circumventing the need for additional DFT data altogether (Appendix~\ref{appendix:spice_idle_baseline}).
When increasing the number of HF labels in SpiceV1->2, we observe the same trends as on QMugsvL: $\Delta$-learning maintains a rather constant offset to the direct learning baseline, the TB energy performs better than PM6 in both IDLe and $\Delta$-learning, and IDLe with TB + SE labels performs best. The same trends hold for the PubChem-Boron-Silicon subset (Appendix~\ref{appendix:boron_silicon}), but due to the large distribution shift, more HF labels are required with all approaches to reach acceptable MAE values.

\paragraph{Larger molecules:}

Extrapolation to larger molecules than the ones in the training set is particularly challenging for NNPs. However, it is particularly relevant given the different scaling cost of LF and HF methods with the molecule size.
In this experiment, we split QMugsvL into a training dataset $A$ of molecules having up to $n_A$ atoms, and an OOD dataset $B$ of molecules having at least $n_B$ atoms. To simulate increasing levels of OOD difficulty, we perform three splits with  $(n_A=60, n_B=61)$, $(n_A=40, n_B=80)$, and $(n_A=30, n_B=120)$.

\begin{wrapfigure}[18]{r}{0.5\textwidth}
    \centering
    \includegraphics[trim= 10 5 30 30, width=0.5\textwidth]{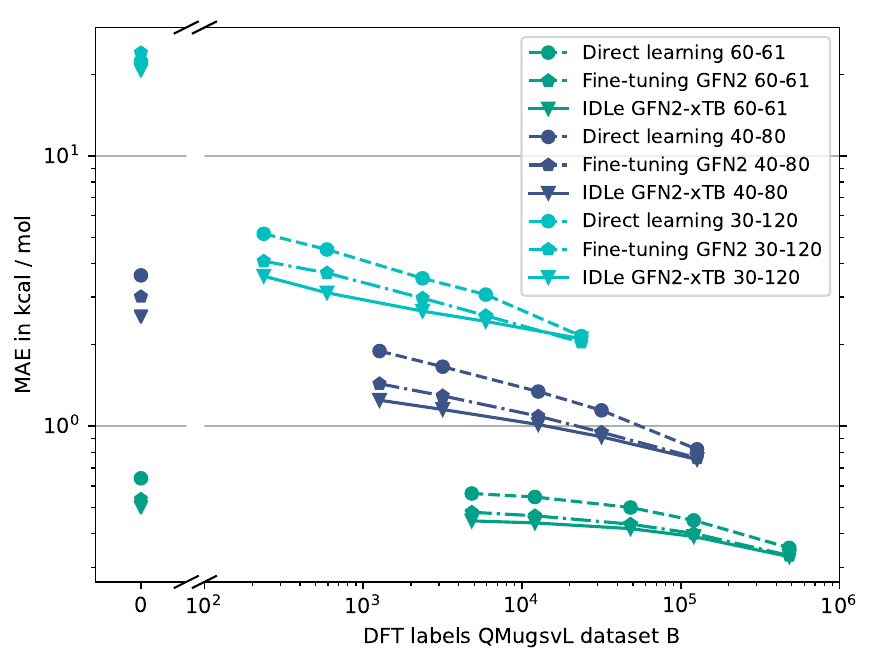}
    \caption{Extrapolation to larger molecules. MAE on datasets $B$ for three different distribution shifts with $(n_A=60, n_B=61)$, $(n_A=40, n_B=80)$ and $(n_A=30, n_B=120)$.}
    \label{fig:qmugs_large}
\end{wrapfigure}

For all splits, IDLe GFN2-xTB outperforms direct learning and fine-tuning for all amounts of DFT labels on the datasets $B$. 
As the distribution shift increases, the need for additional DFT labels increases, as shown by the increasing performance gap between 0\% and 1\% DFT labels (Figure~\ref{fig:qmugs_large}). IDLe leverages the additional HF data most efficiently, reaching almost the same performance as direct learning at 25\% DFT labels while using only 1\% of them.

\subsection{IDLe data scaling and resource gains}
Given the IID and OOD results above, IDLe opens new opportunities to build QM datasets to train HF-NNPs by leveraging LF labels more efficiently.  Therefore, in this section, we study the computational resource gains from leveraging LF labels with IDLe and its scaling behavior with respect to data. 

\paragraph{Computational resource gains}
For a given dataset, let \(N_\mathrm{IDLe}\) represent 1\% of the total available HF labels, and \(N_\mathrm{direct}\) be the amount of HF labels needed by direct learning to reach the same performance as IDLe using only 1\% of the available HF labels. These quantities can be inferred from Figures~\ref{fig:iid_qmugs_qm7x}, \ref{fig:iid_ani1ccx}, and \ref{fig:spice}. Using the time benchmarks in Appendix~\ref{appendix:time_bench}, we estimate the CPU time required to generate all HF labels (\(T_\mathrm{HF}\)) and all LF labels (\(T_\mathrm{LF}\)). Table~\ref{table:comp_pred} shows the ratios of HF data efficiency (\(N_\mathrm{direct} / N_\mathrm{IDLe}\)) and CPU time gains (\(N_\mathrm{direct} \times T_\mathrm{HF} / (0.01 \times T_\mathrm{HF} + T_\mathrm{LF})\)).

These ratios demonstrate that direct learning uses significantly more HF data to achieve the performance that IDLe reaches with only 1\% of the HF data. Factoring in the cost of LF labeling, IDLe shows great promise for reducing CPU time for data generation, particularly when dealing with large molecules and when the HF method is much more expensive (and accurate) than the LF method.

\begin{table}[ht]
\caption{Cost of generating high fidelity (HF) and tight binding (TB) labels for the whole dataset, as well as IDLe data and compute efficiency ratios compared to direct learning.}
\centering
\resizebox{\textwidth}{!}{\begin{tabular}{@{}ccccccc@{}}
\toprule
\multirow{2}{*}{Dataset} & \multicolumn{2}{c}{HF Label}      & \multicolumn{2}{c}{TB Label} & \multicolumn{2}{c}{Ratios at 1\% HF} \\ \cmidrule(l){2-7} 
 &  Method  & CPU Time [d]  & Method   & CPU Time [d]  & Data  & CPU Time [d] \\
 \midrule
 ANI-1ccxvL & CCSD(T)/CBS  &  6.55e+6 & GFN2  & 170  & 25x  & 24.94x \\
SpicevL2  & $\omega$B97M-D3(BJ)/def2-TZVPPD & 6.80e+4 & GFN2  & 728 & 50x  & 24.15x \\ 
QMugsvL   & $\omega$B97X-D/def2-SVP  & 4.36e+4 & GFN2  & 752 & 5x    & 1.84x  \\
QM7-XvL  & PBE0+MBD  & 1.21e+4 & DFT3B & 1487  & 11.5x   & 0.87x\\ 
\bottomrule
\end{tabular}}

\label{table:comp_pred}
\end{table}

\begin{figure}[h]
    \centering
    \includegraphics[trim= 50 5 70 30, clip,width=0.8\textwidth]{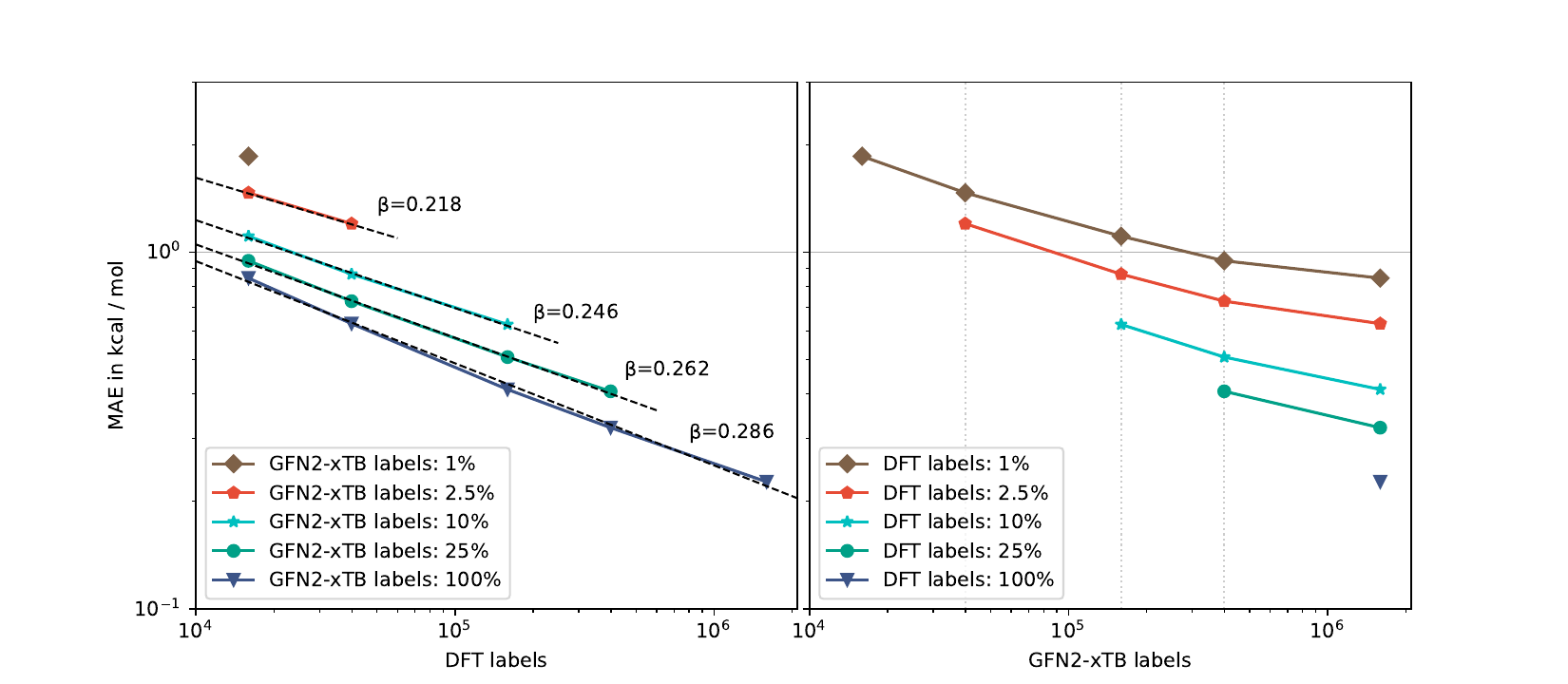}
    \caption{Neural scaling of IDLe. Mean absolute error (MAE) and exponent $\beta$ on QMugsvL when varying the amount of DFT data for a fixed amount of LF GFN2-xTB data (left) and vice versa (right).}
    \label{fig:qmugs_data_scaling}
\end{figure}

\paragraph{IDLe data scaling}
In many ML domains, including NNPs \cite{frey2023neural}, the IID test loss follows an empirical power-law scaling
$\mathcal{L}(R) = \alpha R^{-\beta}$ as a function of resource $R$ (training data, model capacity, compute), where the exponent $\beta$ is the rate of performance improvement \cite{hestness2017deep, kaplan2020scaling, zhai2022scaling}. These power laws tend to break down when the model capacity is insufficient for the data size or vice versa
\cite{bahri2021explaining}. Therefore, we study the scaling laws of IDLe on QMugsvL to determine the limits of increasing the LF or the HF data. We performed this study by fixing the amount of LF data and varying the amount of HF data (left) and vice versa (right) (see Figure~\ref{fig:qmugs_data_scaling}). In the former case, the model perfectly follows power-law scaling when increasing HF data up to the amount of LF data. Importantly, the efficiency (\(\beta\)) towards HF data increases with an increasing amount of LF labels. Conversely, when increasing the amount of LF data beyond the amount of HF data, the model enters a "saturation regime" where \(\beta\) decreases as the amount of LF data increases (Figure~\ref{fig:qmugs_data_scaling}, right). Interestingly, when computing \(\beta\) section-wise (slope of the curve between vertical lines), a secondary trend becomes visible: \(\beta\) increases with an increasing amount of HF labels (Appendix~\ref{table:scaling_betas}).

These findings provide insights into the learning behavior of IDLe: when adding new conformations with LF data, the model performs increasingly well in predicting LF energies. However, we hypothesize that the model's ability to infer HF energies is limited by the size of overlapping representations. Thus, representing LF energies increasingly well exhibits diminishing returns in predicting HF energy (the "saturation regime"). At this point, further increasing the amount of HF labels for existing conformers allows the model to increase the amount of shared representations (up to some limit determined by the specific QM methods involved), enhancing data efficiency for adding more LF conformers.

%% file: sections/discussion.tex
We have demonstrated that IDLe can efficiently leverage energy labels from LF QM methods to reduce the number of required HF QM calculations by up to 50x. This significantly lowers the cost of generating QM datasets for larger molecules and broader chemical space coverage, paving the way for foundational NNPs. We envision next-generation NNPs trained on large multi-fidelity QM datasets to enhance the accuracy of MD simulations, supporting material science and drug discovery.
Several consistent results emerged from our experiments that are worth highlighting:
\begin{itemize}
    \item \textbf{Increased representation power with multi-fidelity}: Training on multi-fidelity labels via IDLe and transfer learning improves the accuracy compared to direct HF learning by a large margin. For IDLe, this even holds for the case of 100\% HF labels, implying more robust representation learning with multi-fidelity labels.
    \item \textbf{Increased OOD Performance with IDLe}: IDLe excels in out-of-distribution (OOD) settings, requiring no additional HF labels for small and medium distribution shifts.
    \item \textbf{Data and training efficiency}: IDLe outperforms $\Delta$-learning in the low data regime and is less sensitive to differences in LF and HF energy values. Unlike $\Delta$-learning, IDLe can improve performance by leveraging multiple LF labels simultaneously. We explain that the enhanced representation capacity is due to the additional LF data. IDLe outperforms or is on par with transfer learning, with IDLe benefiting more strongly from high MI between the LF and HF methods, while transfer learning appears more robust to low MI.
    
\end{itemize}

\textbf{Limitations:}
In this study, all NNPs were trained via energy-matching. However, training on energies + forces
simultaneously is critical to render these models robust and stable during MD simulations in practice. Additionally, for large OOD distribution shifts, all methods, including IDLe, require a significant amount of HF labels.
Nonetheless, IDLe tends to be the most efficient in using these additional HF labels compared to the other methods investigated.

\textbf{Future work:}
To address the above limitations, our future studies will focus on training NNPs on energies, forces and other physical quantities. We will explore multiple HF methods with equal importance and study their scaling laws which can guide resource-efficient data generation. The LF data-scaling and their "saturation regime" are of particular interest, as they differ from the scaling laws of transfer learning \cite{hernandez2021scaling}. Additionally, we will investigate the shared MI between LF and HF methods, providing insights into the NNP latent representations (e.g., GFN2-xTB might have more MI with the $\omega$B97 functionals, while DFTB3 might have less with PBE0+MBD).

%% file: sections/additional-files.tex
\subsection{Hyperparameters}
\label{appendix:hyperparameters}
We employ TorchMDNet2.0 \cite{pelaez2024torchmd} with the same hyperparameters (summarized in table \ref{table:TorchMDNet_hyperparameters}) for all datasets and training approaches throughout this work. The only exception is the learning rate, which we reduced from $4 \cdot 10^{-4}$ to $10^{-4}$ for SpicevL2 for training stability. All training runs were performed on 4 Nvidia A100 GPUs with 8 CPU workers per GPU. The training time of each run varied with the amount of training data, but took less than 5h for ANI1-ccxvL, 1.5 days for QMugsvL, 2.5 days for QM7-XvL, 3 days on SpicevL2.

\begin{table}[h]
\centering
\caption{Hyperparameters used throughout this work}
\begin{tabular}{ll}
\toprule

Hyperparameter name & value \\ \midrule
Initial learning rate (LR) & $3 \cdot 10^{-4}$; $10^{-4}$ (SpicevL2) \\ 
Batch size & 28 \\
Optimizer & Adam \cite{Kingma2015} \\
\midrule
LR Scheduler & ReduceLROnPlateau \\
Decay factor & 0.5 \\
Patience & 10 epochs \\
Minimum LR & $10^{-7}$ \\
\midrule
StochasticWeightAveraging & LR: $10^{-6}$ \\
\midrule
\# Parameters & 1.8 milion \\
Cut-off & 5 A \\
Prediction Heads Hidden Neurons &64 \\
Hidden Channels &128 \\
\# Layers &8 \\
\# Attention Heads &8 \\
Activation Function & Swish \cite{ramachandran2017searching} \\
\# RBF & 64, Trainable \\
Neighbor Embedding & True \\
\bottomrule
\end{tabular}
\label{table:TorchMDNet_hyperparameters}
\end{table}

\subsection{Chemical space of the Spicev2 dataset}
\label{appendix:soap}
We split Spicev2 into Spicev1 and Spicev1->2, which cover different parts of chemical space. In order to visualize the shift between these chemical distributions, we use dscribe \cite{dscribe} to compute Smooth Overlap of Atomic Positions (SOAP) \cite{bartok2013soap} descriptors
and project them onto a 2-dimensional manifold using the Uniform Manifold Approximation and Projection for Dimension Reduction (UMAP)\cite{mcinnes2018umap}.
\begin{figure}[h]
    \centering
    \includegraphics[clip,width=0.4\textwidth]{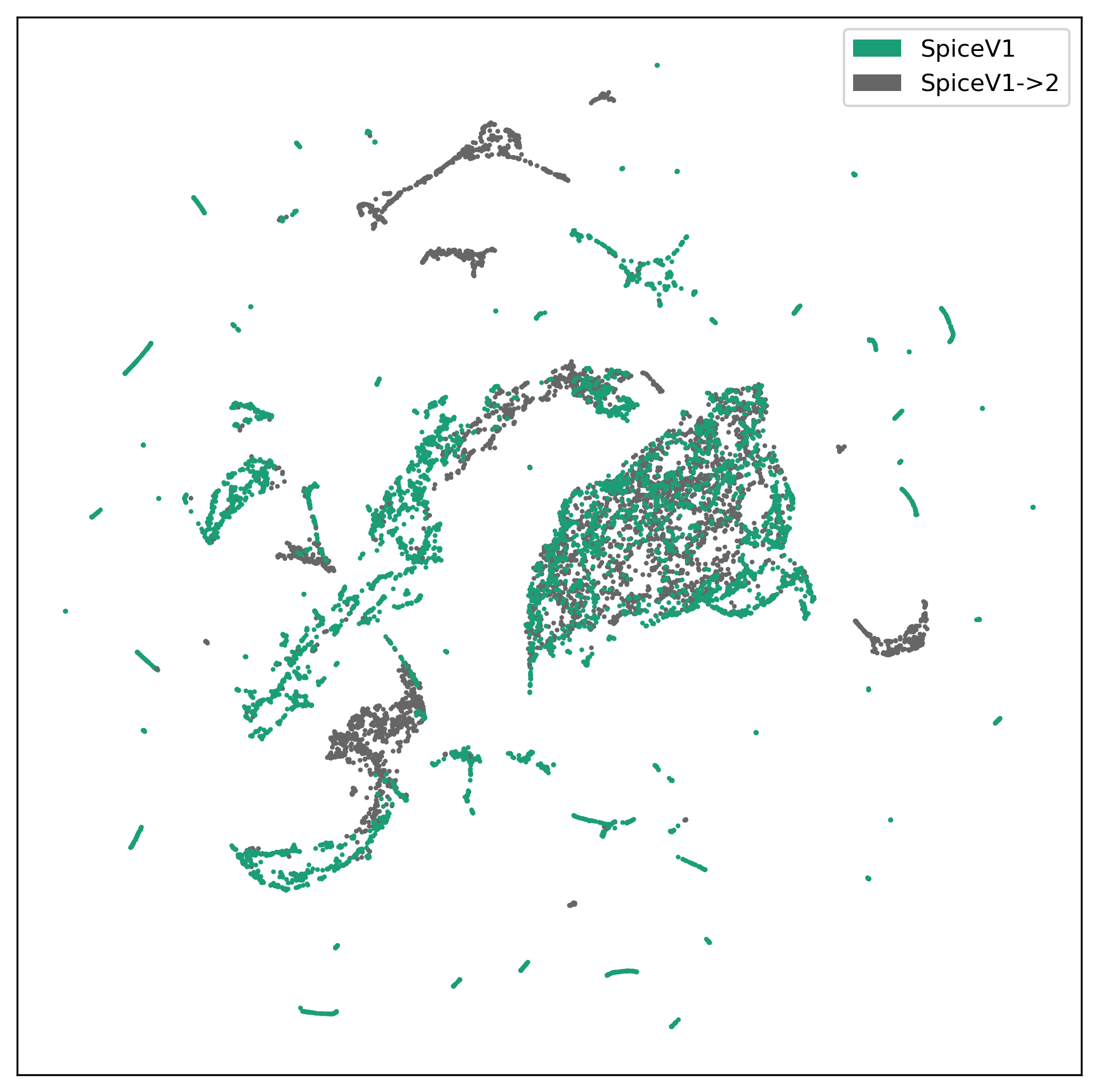}
    \caption{UMAP 2-plot of SOAP descriptors of SpiceV1 and SpiceV1->2}
    \label{fig: SOAP_descriptors_spice}
\end{figure}

\subsection{Performance IDLe verus direct learning for subsets of Spicev1->2}
\label{appendix:spice_idle_baseline}
\begin{figure}[h!]
    \centering
    \includegraphics[trim= 30 5 30 20, clip,width=\textwidth]{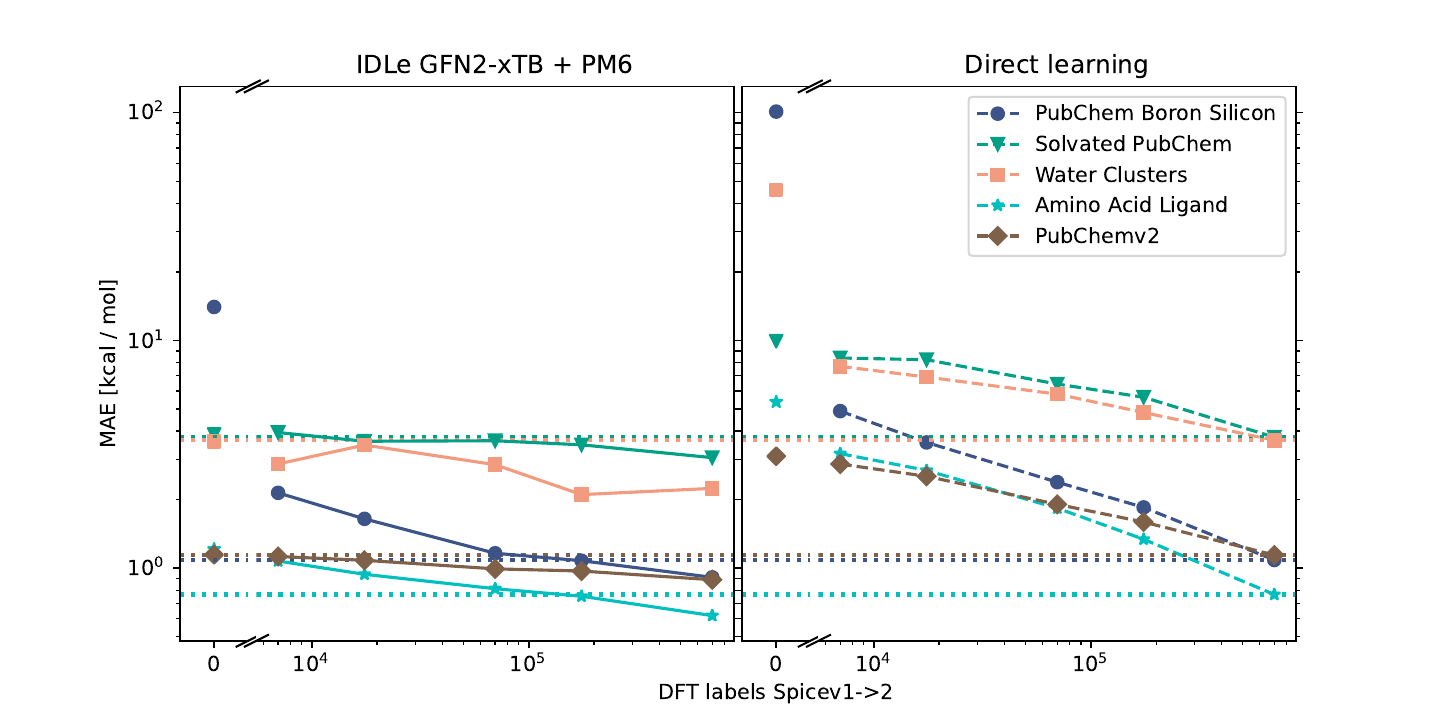}
    \caption{Performance comparison of IDLE GFN2-xTB + PM6 with the direct learning baseline for various subsets of SpiceVL2.}
    \label{fig:spice_idle_vs_baseline}
\end{figure}

Figure \ref{fig:spice_idle_vs_baseline} shows the data-efficiency of IDLE GFN2-xTB + PM6 compared to the direct learning baseline. The PubChemv2 and Water Cluster subsets achieve the same accuracy as direct learning using 100\% of the HF data. Additionally, the error of IDLe on the Solvated PubChem subset is only marginally larger compared to the baseline. On the other hand, the subsets with larger distribution shift require approx. 25\% HF data to reach the 100\% HF baseline accuracy. Hence, when generating new data, it appears to be sufficient to label conformers with small to medium distribution shift via SE methods, saving computational budget for HF computations on molecular structures that are highly dissimilar from existing datasets.

\FloatBarrier

\subsection{PubChem-Boron-Silicon subset performance}
\label{appendix:boron_silicon}
We separate the PubChem-Boron-Silicon Subset of SpicevL2 from the other 4 subsets of Spicev1->2 given that it is extremely out of distribution (Spicev1 containing no B or Si), resulting in large MAE values in the low data regime for all approaches, which would otherwise dominate the results for the remaining subsets. The main difference to Figure \ref{fig:spice} is that all approaches require more data to achieve acceptable accuracy, but the relative performance of the different training approaches is similar.

\begin{figure}[h]
    \centering
    \includegraphics[trim= 15 5 30 20, clip,width=0.6\columnwidth]{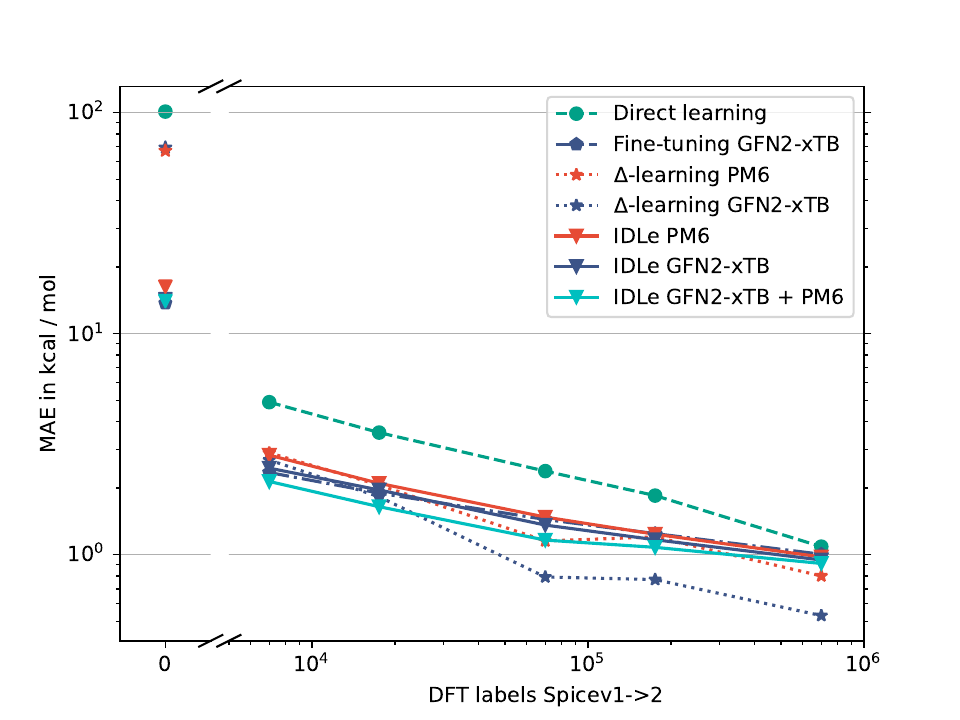}
    \caption{Performance comparison for the PubChem-Boron-Silicon subset of SpicevL2.}
    \label{fig:boron-silicon}
\end{figure}

\FloatBarrier

\subsection{Time benchmarking of various QM methods} \label{appendix:time_bench}



\begin{figure}[h]
    \centering
    \includegraphics[width=0.8\columnwidth]{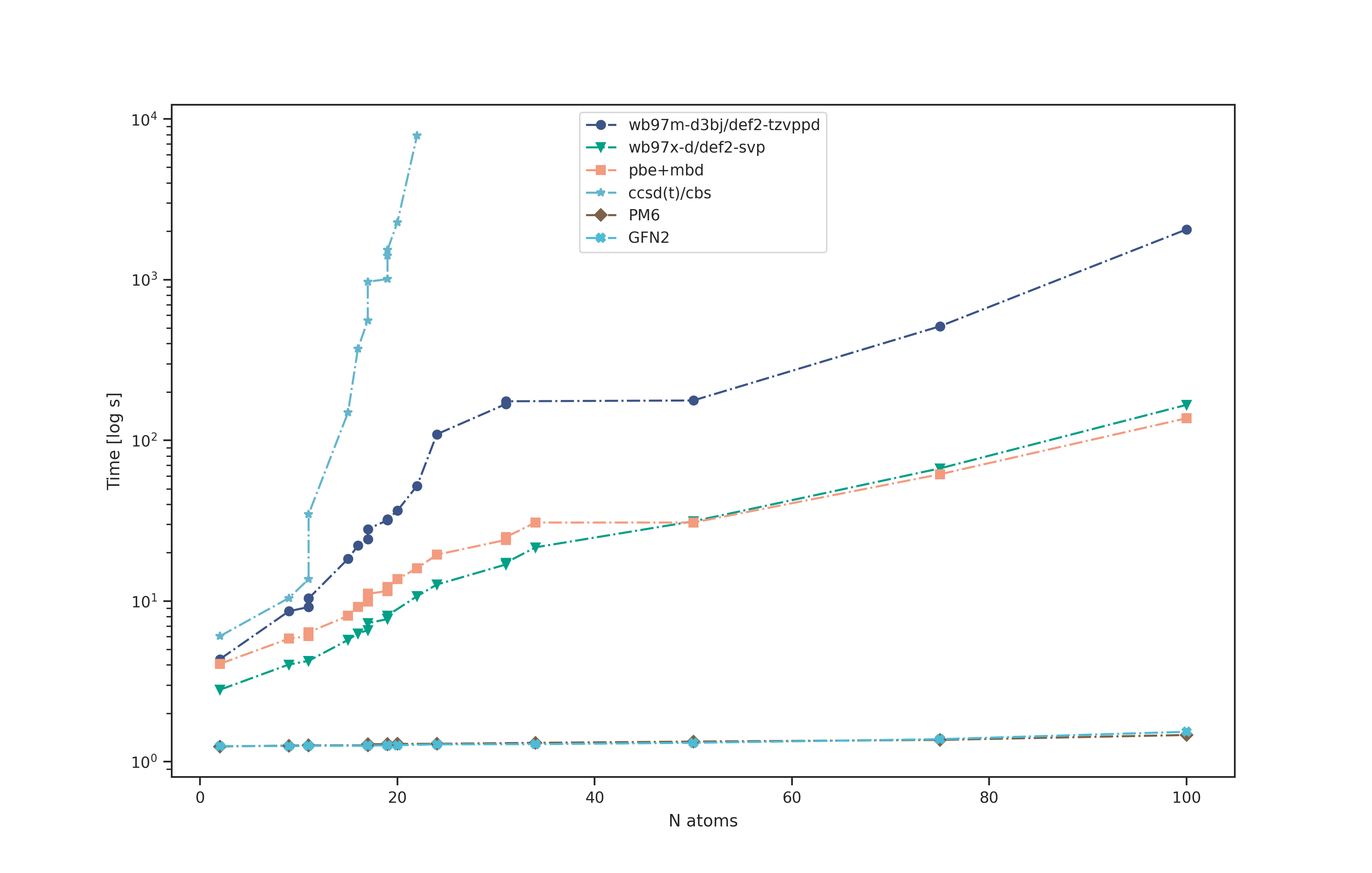}
    \caption{Computational cost of QM computions at various levels of theory scaling with the number of atoms per molecule.}
    \label{fig:ScalingLevelOfTheoriesNatoms}
\end{figure}

We conducted a series of benchmarks on a range of SE, TB, DFT, and CCSD(T) methods to observe the differences in compute time scale required for each method. We selected 17 different small molecules (molecular weights < 250 a.u.), which included some of the most frequently encountered atomic species in the biological field (C, N, O, S, F, Cl) and included three additional small/medium sized solvated system (with molecular weights between 300 and 700 a.u.) to include most of the systems sizes and interaction types covered by the recent \textit{sota} QM datasets. We then correlated the time taken to perform a single point calculation to their exact mass and isotopic molecular weight (Figure~\ref{fig:ScalingLevelOfTheoriesNatoms}) and extrapolated this data to obtain the approximate computation of all the datasets in the study (Table~\ref{table:comp_pred}).

All the benchmarks were executed on an AMD Ryzen Threadripper PRO 3995WX 64-Cores CPU, the calculation were conducted with MOPAC version 22.1.0, 
XTB version 20.2 for the SE single points label; for the electronic quantum mechanical calculation, the software Psi4 \cite{psi42020} version 1.4.1 was used. All the calculations were run on the backend QCEngine version 0.26.0.

In a realistic setting, most of the selected conformations of the datasets are not only larger but also more complex due to the diversity of interactions involved. For instance, SpiceV2 includes additions of bulk water phases and solvated peptides. This increased complexity would ultimately increase the computation time demanded by high-fidelity DFT methods, extending beyond our initial benchmark results. Consequently, the use of SE/TB calculations and IDLe becomes even more appealing due to their demonstrated efficiency. 


\subsection{Data scaling coefficients of IDLe on QMugsvL}
\label{appendix:qmugs_data_scaling}
While the scaling laws of IDLe for increasing HF labels with a fixed amount of LF data follow classical power laws, the scaling behavior when adding conformers with only LF labels is more intricate. Given that the scaling exponent $\beta$ depends both on the total number of SE and HF labels, we compute the slope of the curves in Figure \ref{fig:qmugs_data_scaling} for each section (vertical gray lines; Table \ref{table:scaling_betas}). This analysis reveals two distinct trends: lower data efficiency when increasing the number of LF labels a for fixed number of HF labels and higher data efficiency towards adding more LF labels when increasing the amount of HF labels. Thus, when modeling $\beta$ for scaling LF data, the number of conformers with only LF labels might be more informative than the total number of LF labels.

\begin{table}[h]
\centering
\caption{Segment-wise scaling coefficients $\beta$ on QMugsvL when increasing the percentage of semi-empirical GFN2-xTB labels, for a fixed amount of DFT labels.}
\begin{tabular}{ccccc}
\toprule
\diagbox{DFT labels}{GFN2-xTB labels} & 1\% $\rightarrow$ 2.5\% & 2.5\% $\rightarrow$ 10\%& 10\% $\rightarrow$ 25\% & 25\% $\rightarrow$ 100\% \\ 
 \midrule
 1\% & 0.257 & 0.202 & 0.173 & 0.081 \\
 2.5\% & & 0.235 & 0.190 & 0.105\\ 
  10\% & & & 0.230 & 0.151 \\
 25\% & & & & 0.169 \\ 
\bottomrule
\end{tabular}

\label{table:scaling_betas}
\end{table}